\documentclass[twocolumn,showpacs,preprintnumbers,amsmath,amssymb]{revtex4}

\usepackage{graphicx}
\usepackage{dcolumn}
\usepackage{bm}

\def\OP#1#2#3#4{{\bigl.^{#1}\hspace{-0mm}{#2}_{#3}^{[#4]}}}

\def\be{\begin{equation}}
\def\ee{\end{equation}}
\def\bea{\begin{eqnarray}}
\def\eea{\end{eqnarray}}
\def\NO{\nonumber}
\def\gev{\mathrm{~GeV}}

\def\dfrac{\displaystyle\frac}

\begin{document}


\title{Inclusive Production of $h_c(h_b)$ States via $e^{+}e^{-}$ Annihilation}


\author{Jian-Xiong Wang$^{1,2}$ and Hong-Fei Zhang$^{1,2}$}%
\affiliation{
Institute of High Energy Physics, Chinese Academy Sciences, P.O. Box 918(4), Beijing, 100049, China.\\
Theoretical Physics Center for Science Facilities, CAS, Beijing, 100049, China.
}%
\date{\today}

\begin{abstract}
We calculate the inclusive production of $h_c(h_b)$ at $e^+e^-$ colliders with the center-of-mass energy from the CLEO-c energy 
to $Z^0$ boson mass at leading order of nonrelativistic QCD.  At $Z^0$ boson mass, the cross sections are  $39\sim 703$fb 
for $h_c(1p)$, $37\sim 61$fb for $h_b(1p)$ and $44\sim 73$fb for $h_b(2p)$. At the B-factory, it is $86\sim212$fb for $h_c(1p)$. 
For $h_c$ at the CLEO-c and $h_b(1p,2p)$ at the B-factory, the perturbative QCD expansion is not good and 
the results are much smaller than the experimental measurements. It is clearly shown in all the results that the color-octet
state $(^1S_0^8)$ contributes dominantly while the color-singlet state $(^1P_1^1)$ contributes small or even negative part, in contrast to 
the case of inclusive $J/\psi$ production where the color-singlet state is found to contribute dominantly.
\end{abstract}

\pacs{12.38.-t, 12.39.St, 13.60.Hb, 14.40.Pq}
\maketitle
In recent years, many experimental measurements for P-wave quarkonia $h_c,h_b (1^{+-})$ have been achieved.  
The related branch ratios were measured~\cite{Calderini:2004pd, Andreotti:2005vu, Fang:2006bz, Ablikim:2010rc, Lees:2011zp, Ge:2011kq},
the masses of them are measured precisely~\cite{Rosner:2005ry, Rubin:2005px, Dobbs:2008ec, CLEO:2011aa, Adachi:2011ji},
and the cross sections for $h_c(h_b)$ production via $e^{+}e^{-}$ annihilation at the CLEO-c (B-factory) are also 
measured~\cite{CLEO:2011aa, Adachi:2011ji}. More experimental measurements on P-wave quarkonium states $h_c,h_b$ could be expected 
in the future.  It provides a new place to test or improve our knowledge of quantum chromodynamics (QCD) on 
heavy quarkonium production and decay. 

To study heavy quarkonium decay and production processes, nonrelativistic QCD(NRQCD)~\cite{Bodwin:1994jh} is a 
successful factorization theorem, in which calculation is factorized into  process-dependent short-distance coefficients, 
to be calculated perturbatively in the strong-coupling constant
$\alpha_s$ expansions, and universal long-distance matrix elements (LDMEs), to be extracted from experiment.  
In this scheme, the $Q\overline{Q}$ pair can be produced in any Fock state
$n={^{2S+1}L_J^{[a]}}$ with definite spin $S$, orbital angular momentum
$L$, total angular momentum $J$, and $a=1,8$,
where $a=1$ is for color-singlet (CS) and $a=8$ is for
color-octet (CO) states which finally evolve into physical
quarkonia through nonperturbative processes.
The relative importance of all the CO and CS states are estimated by velocity
scaling rules, which weigh each of the LDMEs by a definite
power of the heavy-quark velocity $v$ in the limit $v\ll1$.
In this way, the theoretical predictions are organized in double expansions in
$\alpha_s$ and $v$.

Based on heavy quark spin symmetry of NRQCD, the LDMEs for $h_c(h_b)$ are simply related to that for $\chi_c(\chi_b)$,
therefore, studies in the past already supply the information of the LDMEs for $h_c(h_b)$ from the fit of $\chi_c(\chi_b)$
related processes.  Experimental data~\cite{Abe:1997yz, LHCb:2012af} shows that $J/\psi$ production from $\chi_c$ feeddown 
count on a large part in 
$J/\psi$ hadroproduction. Based on the experimental data~\cite{Abe:1997yz} and LO calculation on $\chi_c$ hadroproduction,  
an estimate of the CO LDME for $\chi_c$ and $h_c$ was given in Ref.~\cite{Cho:1995vh, Cho:1995ce}. 
Employing these matrix elements, the calculations of $h_c$ hadroproduction at the Tevatron~\cite{Sridhar:1996vd} and 
LHC~\cite{Sridhar:2008sc, Qiao:2009zg} predicted a significant yield.
Photoproduction of $h_c$ was investigated in Ref.~\cite{Fleming:1998md} by using a CO LDME extracted from the decay 
$B\rightarrow\chi_{cJ}+X$, the results indicated a significant cross section at the DESY HERA. 
Recently, the $\chi_c$ hadroproduction calculated up to QCD next-to-leading order (NLO) with an estimate of the CO LDME 
for $\chi_c$ is given in Ref.~\cite{Ma:2010vd}. In the study of polarization for prompt $J/\psi$ hadroproduction~\cite{Gong:2012ug},  
the CO LDME for $\chi_c$ is given by fitting the Tevatron and LHCb data.  

In this paper, we study heavy quarkonium P-wave states $h_c(h_b)$ inclusive production via $e^{+}e^{-}$ annihilation by 
calculating the related cross sections at the leading-order(LO) of NRQCD. Detailed results are given at the $Z^0$ boson peak, B-factory
and CLEO-c. The discussions on the comparison between our calculation and the experimental measurements are given. 

In NRQCD framework, at LO of $\alpha_s$ and $v^2$, cross sections for $H$ state production can be expressed as
\be
\sigma(H)=f_{^{1}P_{1}^{[1]}}\langle O^{H}(^{1}P_{1}^{[1]})\rangle+f_{^{1}S_{0}^{[8]}}\langle O^{H}(^{1}S_{0}^{[8]})\rangle ,
\ee
where $f_n$ denotes the short-distance coefficient corresponding to the NRQCD operator $O^{H}(n)$ and the LDMEs of the operators are
\bea
&&\langle O^{H}(\OP{1}{S}{0}{8})\rangle=\langle 0|\psi^{\dag}T^{a} \chi a_{H}^{\dag}a_{H}\chi^{\dag}T^{a} \psi|0\rangle ,\\
&&\langle O^{H}(\OP{1}{P}{1}{1})\rangle=\langle 0|\psi^{\dag}(-i\dfrac{\overleftrightarrow{\bf D}}{2})
\chi a_{H}^{\dag}\cdot a_{H}\chi^{\dag}(-i\dfrac{\overleftrightarrow{\bm D}}{2})\psi|0\rangle, \NO
\eea
where $H$ represents the hadron state $h_c$, or $c\bar{c}(^1P_1^{[1]},~^1S_0^{[8]})$ states. The CS operator $O^{H}(\OP{1}{P}{1}{1})$
is of order $v^2$ while the CO operator $O^{H}(\OP{1}{S}{0}{8})$ is of order $v^0$.
However, to hadronize into $h_c$, the CO state have to emit at least one soft gluon, 
which rises the order of the LDME $\langle O^{h_c}(\OP{1}{S}{0}{8})\rangle$ 
to $\alpha_{s}v^2$, the same order of $v$ as the CS LDME $\langle O^{h_c}(\OP{1}{P}{1}{1})\rangle$. 

The leading processes for inclusive $h_c$ production are listed as
\bea
&&e^{+}e^{-}\rightarrow c\bar{c}(^{1}P^{\left[1\right]}_{1})+g+g , \NO\\
&&e^{+}e^{-}\rightarrow c\bar{c}(^{1}P^{\left[1\right]}_{1})+c+\bar{c}, \label{eqn:p} \\ 
&&e^{+}e^{-}\rightarrow c\bar{c}(^{1}S^{\left[8\right]}_{0})+g. \NO
\eea
It is easy to see that the three processes are of the same order of $v^2$ at LO. The cross sections for the second and third processes
are infrared (IR) divergence free and can be calculated directly. The cross section for the first process, although at LO, 
is IR divergent when one of the gluons is soft. To deal with the divergence, the NRQCD factorization formulas involved are as follows,
\be
\sigma(h_c)=f_{^1P_1^{[1]}}\langle O^{h_c}(P_1^{[1]})\rangle
+f_{^{1}S^{[8]}_{0}}\langle O^{h_c}(^1S_0^{[8]})\rangle 
\label{eqn:s}
\ee 
and 
\bea
&\sigma(^{1}P^{[1]}_{1})=f_{^1P_1^{[1]}}\langle O^{^1P_1^{[1]}}(^{1}P_1^{[1]})\rangle
+f_{^1S^{[8]}_{0}}\langle O^{^1P_1^{[1]}}(^1S_0^{[8]})\rangle , \NO \\
&\sigma(^1S_0^{[8]})=f_{^1P_1^{[1]}}\langle O^{^1S_0^{[8]}}(^1P_1^{[1]})\rangle
+f_{^{1}S^{[8]}_{0}}\langle O^{^1S_0^{[8]}}(^1S_0^{[8]})\rangle.  
\label{eqn:sig}
\eea
The matrix elements of operators are calculated in dimensional regularization scheme at LO with an ultraviolet cutoff (NRQCD scale) 
$\mu_{\Lambda}=m_{c}$ and are given as:
\bea
&\langle O^{^1S_0^{[8]}}(^1P_1^{[1]})\rangle=0 , \label{eqn:LDM} \\  
&\langle O^{^1P_1^{[1]}}(^{1}S_{0}^{[8]})\rangle=-\frac{\alpha_s}{3\pi m_{c}^2}u^{c}_{\epsilon}\frac{N_{c}^{2}-1}{N_{c}^{2}}
\langle O^{^1P_1^{[1]}}(^{1}P_{1}^{[1]})\rangle , \NO
\eea
where $N_c$ is 3 for $SU(3)$ gauge field and $u_\epsilon$ is defined as 
\be
u^{c}_{\epsilon}=\frac{1}{\epsilon_{IR}}-\gamma_{E}+\frac{5}{3}+ln(\frac{\pi\mu_{R}^{2}}{\mu_{\Lambda}^2}) 
\ee
with $\mu_R$ being the renormalization scale.
The divergence in this LDME will cancel that in $\sigma(^{1}P^{\left[1\right]}_{1})$,
which can be isolated by using the two-cutoff phase space slicing method~\cite{Harris:2001sx} as
\be
\sigma(^{1}P^{\left[1\right]}_{1})=\sigma^{S}(\OP{1}{P}{1}{1})+\sigma^{\bar{S}}(\OP{1}{P}{1}{1}) ,
\ee
where $\sigma^{\bar{S}}$ and $\sigma^{S}$ are from the gluon-soft and hard region, respectively.
The boundary of the two regions is $E_{g}=\frac{\sqrt{s}}{2} \delta_s$,
where $E_g$ is the energy of the soft gluon, $\sqrt{s}$ is the center-of-mass (CM) energy of $e^+e^-$ colliders 
and $\delta_s$ is an arbitrary positive number small enough to make 
the soft approximation good enough.  The soft part can be expressed as
\be
\sigma^{\bar{S}}(\OP{1}{P}{1}{1})=-\frac{\alpha_{s}}{3\pi m_{c}^{2}}u_{\epsilon}^{s}\frac{N_{c}^{2}-1}{N_{c}^{2}}f_{^{1}S_{0}^{[8]}} ,
\ee
where
\be
u_{\epsilon}^{s}=\frac{1}{\epsilon_{IR}}+\frac{s+4m_{c}^{2}}{s-4m_{c}^{2}}ln(\frac{s}{4m_{c}^{2}})+ln(\frac{4\pi\mu^{2}_{R}}{s\delta_{s}^{2}})-\gamma_{E}-\frac{1}{3} .
\ee
Using Eq.(\ref{eqn:sig}) and  Eq.(\ref{eqn:LDM}), we obtain the expressions of short-distance coefficients
in Eq.(\ref{eqn:s}) as
\bea
&f_{^{1}S^{[8]}_{0}}=\frac{\sigma(^1S_0^{[8]})}{\langle O^{^1S_0^{[8]}}(^1S_0^{[8]})\rangle} ,\\  
&f_{^1P_1^{[1]}}=-\frac{\alpha_{s}}{3\pi m_{c}^{2}}\frac{N_{c}^{2}-1}{N_{c}^{2}}u_{\epsilon}f_{^1S_0^{[8]}}
+\frac{\sigma^{\bar{S}}(^1P_1^{[1]})}{\langle O^{^1P_1^{[1]}}(^1P_1^{[1]})\rangle} ,
\label{eqn:soft}
\eea
where
\be
u_{\epsilon}=u_{\epsilon}^{s}-u_{\epsilon}^{c}=\frac{s+4m_{c}^{2}}{s-4m_{c}^{2}}ln(\frac{s}{4m_{c}^{2}})-ln(\frac{s\delta_{s}^{2}}{4\mu_{\Lambda}^{2}})-2.
\ee

It is clearly shown that all the short-distance coefficients are IR divergence free and the cross section for the first process
is well defined. To calculate $\sigma(^1S_0^{[8]})$ and $\sigma^{S}(^1P_1^{[1]})$ for the first process and the cross sections
for the second and third processes, we apply our Feynman Diagram Calculation package(FDC)~\cite{Wang:2004du} to generate all the 
needed FORTRAN source.

From the heavy quark spin symmetry of the NRQCD Lagrangian, it is obvious that 
the LDME $\langle O^{h_{c}}(^1S_0^{[8]})\rangle$ for the intermediate state 
$c\bar{c}(^{1}S^{\left[8\right]}_{0})$ evolving into $h_c$ is exactly the same as 
that for the intermediate state $c\bar{c}(^3S_1^{[8]})$ evolving into $\chi_c$ at LO of $v^2$.
It gives 
\be
\langle O^{h_{c}}(^1S_0^{[8]})\rangle\approx\langle O^{\chi_{c1}}(^3S_1^{[8]})\rangle. 
\ee
We will employ the LDME value obtained in Ref.~\cite{Cho:1995ce} as $\langle O^{\chi_{c1}}(^3S_1^{[8]})\rangle=0.0098\gev^3$,
which is from a fit of the Tevatron experimental data with the LO calculation for transverse momentum $p_t$ distribution of 
$\chi_{cJ}$ hadroproduction and is independent on the NRQCD scale $\mu_\Lambda$ since the range of $p_t$ in the fit does not include 
$p_t=0$ where $\mu_\Lambda$ is involved in the calculation to factorize IR divergence.   
However, the result in our calculation is $\mu_\Lambda$ dependent, therefore we have to fix a value of $\mu_\Lambda$.
Since $\mu_\Lambda$-dependent term is proportional to $ln(\mu_{\Lambda})$ and it is found that different choices of $\mu_\Lambda$ in a 
reasonable range around the heavy quark mass $m_Q$ do not cause larger uncertainty in final results when the CM energy is far above $2m_Q$,
the default value $\mu_{\Lambda}=m_{Q}$ is chosen for $h_c$ production at both B-factories and Z-factory, and for $h_b$ production at the Z-factory.
For $h_c$ production at the CLEO-c and $h_b$ production at B-factories, we give results for different $\mu_\Lambda$ choices to show the uncertainties caused by this parameter.
In the numerical calculation, we have the following common choices as 
$\langle O^{h_c}(P_1^{[1]})\rangle=\frac{27}{2\pi}|R'_{h_c}(0)|^{2}=0.322\gev^5$, $\alpha=1/128$, $\sin^2\theta_W=0.226$, $m_z=91.2\gev$ and $\Gamma_z=2.49\gev$. 
And other parameters are explicitly given in each individual case. 
The soft cutoff $\delta_s$ independence is checked in the calculation and $\delta_s=0.0001$ is used. 

For the CLEO-c experiment at the CM energy $\sqrt{s}=4.17$GeV, 
only the first and last processes in Eq.(\ref{eqn:p}) contribute
and the gluons in the first process are too soft that the convergence of perturbative calculation in QCD is not good.
Therefore, it is not surprising to see that the theoretical results can not describe the experimental measurement.
To show the uncertainty from the c-quark mass $m_c$, the QCD coupling constant $\alpha_s$ and NRQCD scale $\mu_\Lambda$, 
our results are listed in Table.~\ref{table:h_c_c} by choosing $m_c$ as $1.3\gev$, $1.5\gev$ and $1.76\gev$,
$\alpha_s$ as $\alpha_s(3.0\gev)=0.26$ and $\alpha_s(\sqrt{s}/2)=0.30$, and $\mu_\Lambda$ as $m_c$ and $m_c/2$. 

\begin{table}[htbp]
\begin{center}
\begin{tabular}{|c c|}
\hline
$\alpha_{s}(2m_{c})=0.26$&
\begin{tabular}{|c|c|c|c|}
$m_c(\gev)$&$\OP{1}{P}{1}{1}gg$&$\OP{1}{S}{0}{8}g$&Total\\
\hline
1.3&-4.449(-1.534)&4.406&-0.04(2.87)\\
\hline
1.5&-3.057(-1.560)&3.014&-0.04(1.45)\\
\hline
1.76&-1.656(-1.104)&1.531&-0.12(0.43)\\
\end{tabular}\\
\hline
$\alpha_{s}(\sqrt{s}/2)=0.30$&
\begin{tabular}{|c|c|c|c|}
$m_c(\gev)$&$\OP{1}{P}{1}{1}gg$&$\OP{1}{S}{0}{8}g$&Total\\
\hline
1.3&-5.923(-2.042)&5.084&-0.84(3.04)\\
\hline
1.5&-4.071(-2.077)&3.478&-0.59(1.40)\\
\hline
1.76&-2.204(-1.470)&1.766&-0.44(0.30)\\
\end{tabular}\\
\hline
\end{tabular}
\caption{Total Cross sections(pb) for $h_c$ production from $\OP{1}{P}{1}{1}$, $\OP{1}{S}{0}{8}$ and the sum of them with different 
$m_c$, $\alpha_s$ and $\mu_\Lambda$ at $\sqrt{s}=4.170\gev$.
Values in the brackets correspond to $\mu_{\Lambda}=m_{c}/2$.}
\label{table:h_c_c}
\end{center}
\end{table}

The results in the table are all negative for $\mu_\Lambda=m_c$ and become positive for $\mu_\Lambda=m_c/2$ 
while the experimental measurements at the CLEO-c~\cite{CLEO:2011aa} 
give the cross section for $e^+e^-\rightarrow h_c(1P)+\pi^+\pi^-$ as $15.6\pm2.3\pm1.9\pm3.0$pb.  
On the results, there are three points to be addressed. 

First, we focus on the $\alpha_s$ expansion. The value of $\alpha_s$ depends on the renormalization scale $\mu_R$,
and in the total cross section, there are logarithm terms $ln(E_{s}/\mu_{R})$ 
which must be small to make sure that they do not ruin the expansion.
$E_s$ can be $m_c$, $s$ and $p_{max}^g=(s-4m_{c}^2)/(2\sqrt{s})$. 
The smallest scale $p_{max}^g$ is $1.0(0.6)\gev$ when $m_c$ is $1.5\gev(m_{h_c}/2)$, as a result, there is not a proper
choice of $\mu_R$ to lead to good convergence in perturbative QCD expansion. 
The perturbative expansion becomes better when the CM energy is larger than $\sqrt{s}=6\gev$ where $p_{max}^g$ is $2\gev$. 
For inclusive $h_b$ production at the B-factory, it becomes better when the CM energy is larger than $12\gev$ 
where $p_{max}^g$ is $1.8\gev$. 

Second, we investigate the $v^2$ expansion in shot-distance coefficients. 
The worst propagators for the $v^2$ expansion in the first process in Eq.(\ref{eqn:p}) is
\bea
&\frac{1}{(\frac{p_{H}}{2}+q-p_{g})^{2}-m_{c}^{2}}=-\frac{1}{p_{H}\cdot p_{g}}\frac{1-a}{1-a^2}, 
~a=\frac{2 p_{g}\cdot q}{p_{H}\cdot p_{g}}   
\eea
where $p_H$, $p_g$ and $q$ denote the momentum of $h_c$, gluon and the relative momentum of c-quarks in $h_c$, respectively.
It is easy to obtain $a$, a Lorentz invariant, as $a=|q|cos\theta_{q,p_g}/m_c$ in $h_c$ rest frame where $|q|/m_c$ is 
the velocity $v$ of c-quark in the quarkonium. Therefore the $a^2$ expansion in the propagator is the nonrelativistic expansion. 
It is easy to see that validity of the expansion only depends on the value of $a^2$ and has no relation with the momentum of the gluons.
In our case of $h_c$, $v^2$ is thought to be larger than that of $J/\psi$, so the expansion is better for $J/\psi$ than that for $h_c$. 

Third, we have to check more detail of NRQCD factorization scheme.
The default NRQCD scale $\mu_\Lambda=m_c$ sets up a boundary between perturbative and non-perturbative parts.
The maximum energy of the soft gluons in the non-perturbative part is $\mu_\Lambda=m_c$ which even exceeds the maximum
gluon energy $p_{max}^g=1\gev$ in the phase space in the $e^+e^-\rightarrow h_c(1P)+gg$ at CLEO-c energy.
So $\mu_\Lambda=m_c$ is not a suitable choice, and $\mu_\Lambda=m_c/2$, which is smaller than $p_{max}^g$, 
is chosen to present our results. 
Finally it is clear that the uncertainty from the choice of NRQCD scale $\mu_\Lambda$ is larger near production threshold
and with choice of $\mu_\Lambda=m_c/2$, results for $h_c$ inclusive
production at the CLEO-c is $0.30\sim3.04$pb in our calculation which is just $1/50\sim1/5$ of the measurement of the exclusive process
$e^+e^-\rightarrow h_c(1P)+\pi^+\pi^-$.  

\begin{figure}
\center{
\includegraphics*[scale=0.4]{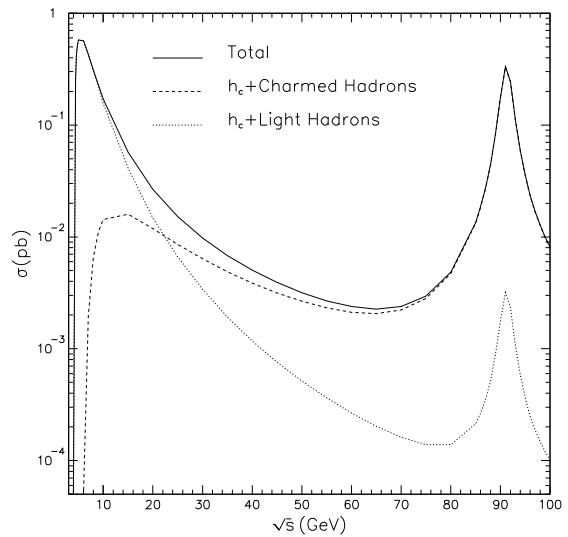}
\caption {\label{fig:hcs} total cross section for production of $h_c$ via $e^{+}e^{-}$ annihilation as a function of CM energy. The parameters are chosen as $m_c=1.5\gev$ and $\alpha_{s}=0.26$.
}}
\end{figure}

In Fig.~\ref{fig:hcs}, we present the dependence of the total cross section for the inclusive $h_c$ production on the CM energy
with $m_c=1.5\gev$ and $\alpha_s=0.26$.
When the CM energy is smaller than $6\gev$, the perturbative QCD expansion is not good and 
the total cross section turns out to be sensitive to the choice of NRQCD scale
$\mu_\Lambda$ and is of large uncertainty. The uncertainty in our result becomes smaller quickly as the CM energy goes larger.
We find that there are two peaks in the plot, one is at $Z^0$ boson mass and the other is at about $5\gev$. And the ratio of 
the cross sections
at the two peaks is about 1.1:2, which means that they are in the same order of magnitude. It can also be seen that the  
$e^+e^-\rightarrow h_c(1P)+gg$ and $e^+e^-\rightarrow h_c(^1S_0^{[8]})+g$ dominate the inclusive production rate when the CM energy 
is smaller than $20\gev$, while $e^+e^-\rightarrow h_c(1P)+cc$ dominate that when the CM energy is larger than $20\gev$. 

All the above discussions are applicable to $h_b$ production at the B-factory with better convergence for
nonrelativistic expansion since $v^2$ becomes smaller. In the numerical calculation, we employ 
the LDMEs, of which the CO ones~\cite{Cho:1995vh} are estimated based on the NRQCD scaling rule and 
the CS ones~\cite{Eichten:1995ch} are obtained based on potential model, as $\langle O^{h_b}(1P)(P_1^{[1]})\rangle=6.1\gev^5$,  
$\langle O^{h_b}(2P)(P_1^{[1]})\rangle=7.1\gev^5$,  
$\langle O^{h_{b}(1P)}(^1S_0^{[8]})\rangle=0.43\gev^3$ and $\langle O^{h_{b}}(2P)(^1S_0^{[8]})\rangle=0.52\gev^3$. 
In Table.~\ref{table:h_b_1} and ~\ref{table:h_b_2}, the results for $h_b(1P)$ and $h_b(2P)$ production at the B-factory are listed,
where $\mu_\Lambda=m_b/4$ is chosen with the same consideration as for $h_c$ case. 
We can see that the uncertainty from the choice of NRQCD scale $\mu_\Lambda$ becomes smaller than that in $h_c$ production at the CLEO-c.  
In comparison with the experimental measurement in Ref.~\cite{Adachi:2011ji}, where exclusive production rate of 
$e^{+}e^{-}\rightarrow h_{b}\pi^{+}\pi^{-}$ are $0.416$pb and $0.695$pb for $h_b(1P)$ and $h_b(2P)$, respectively, our results 
are only $1/13\sim1/5$ and $1/19\sim1/7$ for $h_b(1P)$ and $h_b(2P)$, respectively.

\begin{table}[htbp]
\begin{center}
\begin{tabular}{|c c|}
\hline
$\alpha_{s}(2m_{b})=0.18$&
\begin{tabular}{|c|c|c|c|}
$m_b(\gev)$&$\OP{1}{P}{1}{1}gg$&$\OP{1}{S}{0}{8}g$&Total\\
\hline
4.75&-3.77(-1.50)&76.7&73(75)\\
\hline
4.95&-2.78(-1.34)&53.1&50(52)\\
\hline
5.13&-1.88(-1.05)&32.7&31(32)\\
\end{tabular}\\
\hline
$\alpha_{s}(\sqrt{s}/2)=0.21$&
\begin{tabular}{|c|c|c|c|}
$m_b(\gev)$&$\OP{1}{P}{1}{1}gg$&$\OP{1}{S}{0}{8}g$&Total\\
\hline
4.75&-5.13(-2.04)&89.5&84(87)\\
\hline
4.95&-3.78(-1.82)&62.0&58(60)\\
\hline
5.13&-2.56(-1.43)&38.2&36(37)\\
\end{tabular}\\
\hline
\end{tabular}
\caption{Total Cross sections(fb) for $h_b(1P)$ production from $\OP{1}{P}{1}{1}$, $\OP{1}{S}{0}{8}$ and the sum of them with different $m_b$, $\alpha_s$ and $\mu_\Lambda$, at $\sqrt{s}=10.865\gev$.
Values in the brackets correspond to $\mu_{\Lambda}=m_{b}/4$.}
\label{table:h_b_1}
\end{center}
\end{table}

\begin{table}[htbp]
\begin{center}
\begin{tabular}{|c c|}
\hline
$\alpha_{s}(2m_{b})=0.18$&
\begin{tabular}{|c|c|c|c|}
$m_b(\gev)$&$\OP{1}{P}{1}{1}gg$&$\OP{1}{S}{0}{8}g$&\ \ \ Total\ \ \  \\
\hline
4.75&-4.39(-1.75)&92.8&88(91)\\
\hline
4.95&-3.25(-1.56)&64.2&61(63)\\
\hline
5.13&-2.19(-1.22)&39.5&37(38)\\
\end{tabular}\\
\hline
$\alpha_{s}(\sqrt{s}/2)=0.21$&
\begin{tabular}{|c|c|c|c|}
$m_b(\gev)$&$\OP{1}{P}{1}{1}gg$&$\OP{1}{S}{0}{8}g$&\ \ \ Total\ \ \  \\
\hline
4.75&-5.98(-2.38)&108.3&102(106)\\
\hline
4.95&-4.42(-2.12)&74.9&70(73)\\
\hline
5.13&-2.98(-1.67)&46.1&43(44)\\
\end{tabular}\\
\hline
\end{tabular}
\caption{Total Cross sections(fb) for $h_b(2P)$ production from $\OP{1}{P}{1}{1}$, $\OP{1}{S}{0}{8}$ and the sum of them with different $m_b$, $\alpha_s$ and $\mu_\Lambda$, at $\sqrt{s}=10.865\gev$.
Values in the brackets correspond to $\mu_{\Lambda}=m_{b}/4$.}
\label{table:h_b_2}
\end{center}
\end{table}

\begin{figure}
\center{
\includegraphics*[scale=0.4]{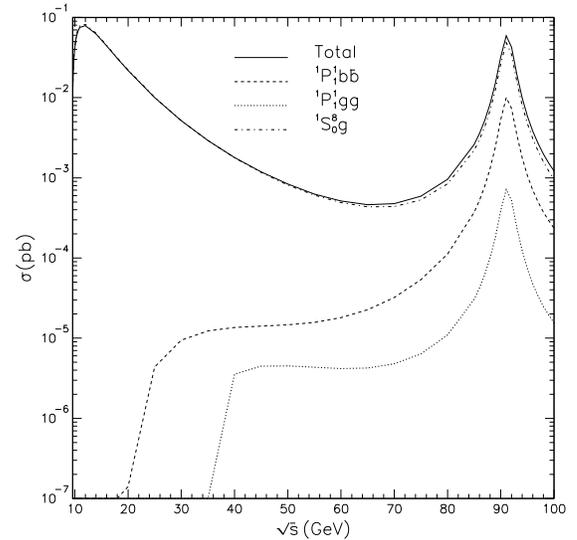}
\caption {\label{fig:hbs} total cross section for production of $h_b$ via $e^{+}e^{-}$ annihilation as a function of CM energy. The parameters are chosen as $m_b=4.75\gev$ and $\alpha_{s}=0.18$.
}}
\end{figure}

Fig.~\ref{fig:hbs} shows the dependence of the total cross section for inclusive $h_b(1P)$ production on the CM energy.
We can see that the $\OP{1}{S}{0}{8}$ process dominates throughout the $\sqrt{s}$ range,
and the first peak where the cross section reaches its maximum value is at the CM energy $12\gev$,
while $Z^0$ boson mass is the second peak.
And the ratio of the cross sections at the two peaks is about 4:3,
which means they are almost of the same value. 

We calculate the production of $h_c$ at the B-factory by changing quark mass and QCD coupling constant to show the uncertainties caused 
by different choices of the two parameters.  The results are listed in Table.~\ref{table:h_c_b}.
In this case, theoretical predictions should be better than those at the CLEO-c since the uncertainty from the choice of $\mu_\Lambda$
becomes much smaller.

\begin{table}[htbp]
\begin{center}
\begin{tabular}{|c c|}
\hline
$\alpha_{s}(2m_{c})=0.26$&
\begin{tabular}{|c|c|c|c|c|}
$m_c(\gev)$&$\OP{1}{P}{1}{1}gg$&$\OP{1}{P}{1}{1}c\bar{c}$&$\OP{1}{S}{0}{8}g$&Total\\
\hline
1.3&8.4&46.2&162.3&212\\
\hline
1.5&-6.5&15.5&137.7&147\\
\hline
1.76&-12.2&3.7&113.5&105\\
\end{tabular}\\
\hline
$\alpha_{s}(\sqrt{s}/2)=0.21$&
\begin{tabular}{|c|c|c|c|c|}
$m_c(\gev)$&$\OP{1}{P}{1}{1}gg$&$\OP{1}{P}{1}{1}c\bar{c}$&$\OP{1}{S}{0}{8}g$&Total\\
\hline
1.3&5.5&30.2&131.1&167\\
\hline
1.5&-4.2&10.1&111.2&117\\
\hline
1.76&-7.9&2.4&91.7&86\\
\end{tabular}\\
\hline
\end{tabular}
\caption{Total Cross sections(fb) for $h_c$ production from $\OP{1}{P}{1}{1}$, $\OP{1}{S}{0}{8}$ and the sum of them with different $m_c$ and $\alpha_s$, at $\sqrt{s}=10.6\gev$.}
\label{table:h_c_b}
\end{center}
\end{table}

The results at $Z^0$ boson mass are listed in Table.~\ref{table:z_c}, Table.~\ref{table:z_b1} and Table.~\ref{table:z_b2} 
for $h_c(1P)$, $h_b(1P)$ and $h_b(2P)$ respectively.   

\begin{table}[htbp]
\begin{center}
\begin{tabular}{|c c|}
\hline
$\alpha_{s}(2m_{c})=0.26$&
\begin{tabular}{|c|c|c|c|c|}
$m_c(\gev)$&$\OP{1}{P}{1}{1}gg$&$\OP{1}{P}{1}{1}c\bar{c}$&$\OP{1}{S}{0}{8}g$&Total\\
\hline
1.3&2.51&698&2.01&703\\
\hline
1.5&1.53&339&1.74&342\\
\hline
1.76&0.89&151&1.48&153\\
\end{tabular}\\
\hline
$\alpha_{s}(\sqrt{s}/2)=0.13$&
\begin{tabular}{|c|c|c|c|c|}
$m_c(\gev)$&$\OP{1}{P}{1}{1}gg$&$\OP{1}{P}{1}{1}c\bar{c}$&$\OP{1}{S}{0}{8}g$&Total\\
\hline
1.3&0.63&175&1.01&177\\
\hline
1.5&0.38&85&0.87&86\\
\hline
1.76&0.22&38&0.74&39\\
\end{tabular}\\
\hline
\end{tabular}
\caption{Total Cross sections(fb) for $h_c$ production from $\OP{1}{P}{1}{1}$, $\OP{1}{S}{0}{8}$ and the sum of them with different $m_c$ and $\alpha_s$, at $\sqrt{s}=91.2\gev$.}
\label{table:z_c}
\end{center}
\end{table}

\begin{table}[htbp]
\begin{center}
\begin{tabular}{|c c|}
\hline
$\alpha_{s}(2m_{b})=0.18$&
\begin{tabular}{|c|c|c|c|c|}
$m_b(\gev)$&$\OP{1}{P}{1}{1}gg$&$\OP{1}{P}{1}{1}b\bar{\bar{b}}$&$\OP{1}{S}{0}{8}g$&Total\\
\hline
4.75&0.73&10.3&49.6&61\\
\hline
4.95&0.63&8.24&47.5&56\\
\hline
5.13&0.54&6.79&45.8&53\\
\end{tabular}\\
\hline
$\alpha_{s}(\sqrt{s}/2)=0.13$&
\begin{tabular}{|c|c|c|c|c|}
$m_b(\gev)$&$\OP{1}{P}{1}{1}gg$&$\OP{1}{P}{1}{1}b\bar{\bar{b}}$&$\OP{1}{S}{0}{8}g$&Total\\
\hline
4.75&0.38&5.37&35.8&42\\
\hline
4.95&0.33&4.30&34.3&39\\
\hline
5.13&0.28&3.54&33.1&37\\
\end{tabular}\\
\hline
\end{tabular}
\caption{Total Cross sections(fb) for $h_b(1P)$ production from $\OP{1}{P}{1}{1}$, $\OP{1}{S}{0}{8}$ and the sum of them with different $m_b$ and $\alpha_s$, at $\sqrt{s}=91.2\gev$.}
\label{table:z_b1}
\end{center}
\end{table}

\begin{table}[htbp]
\begin{center}
\begin{tabular}{|c c|}
\hline
$\alpha_{s}(2m_{b})=0.18$&
\begin{tabular}{|c|c|c|c|c|}
$m_b(\gev)$&$\OP{1}{P}{1}{1}gg$&$\OP{1}{P}{1}{1}b\bar{\bar{b}}$&$\OP{1}{S}{0}{8}g$&Total\\
\hline
4.75&0.85&12.0&60.0&73\\
\hline
4.95&0.73&9.61&57.4&68\\
\hline
5.13&0.63&7.92&55.4&64\\
\end{tabular}\\
\hline
$\alpha_{s}(\sqrt{s}/2)=0.13$&
\begin{tabular}{|c|c|c|c|c|}
$m_b(\gev)$&$\OP{1}{P}{1}{1}gg$&$\OP{1}{P}{1}{1}b\bar{\bar{b}}$&$\OP{1}{S}{0}{8}g$&Total\\
\hline
4.75&0.44&6.26&43.3&50\\
\hline
4.95&0.38&5.02&41.5&47\\
\hline
5.13&0.33&4.13&40.0&44\\
\end{tabular}\\
\hline
\end{tabular}
\caption{Total Cross sections(fb) for $h_b(2P)$ production from $\OP{1}{P}{1}{1}$, $\OP{1}{S}{0}{8}$ and the sum of them with different $m_b$ and $\alpha_s$, at $\sqrt{s}=91.2\gev$.}
\label{table:z_b2}
\end{center}
\end{table}

\begin{figure}
\center{
\includegraphics*[scale=0.4]{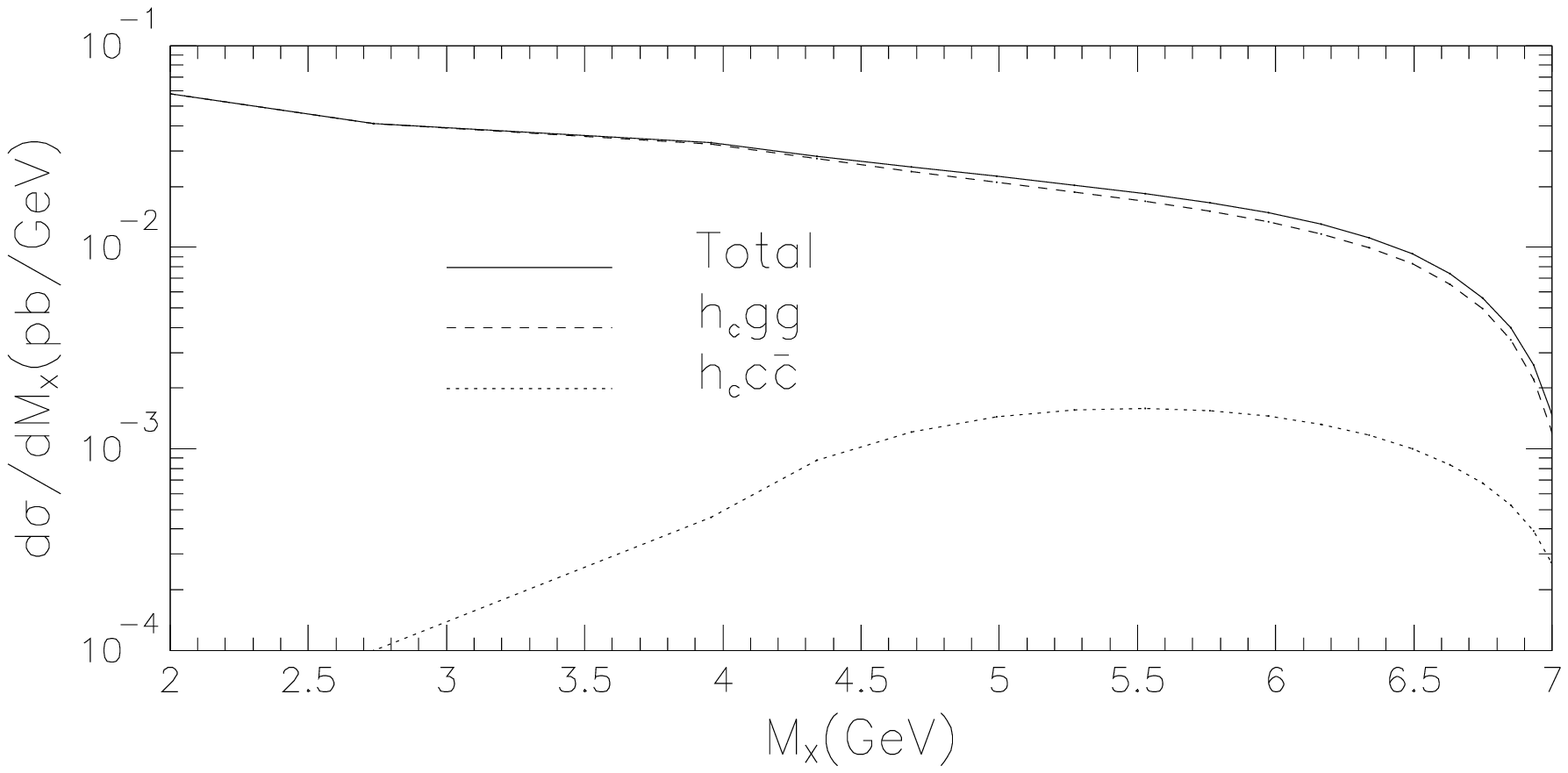}\\
\includegraphics*[scale=0.4]{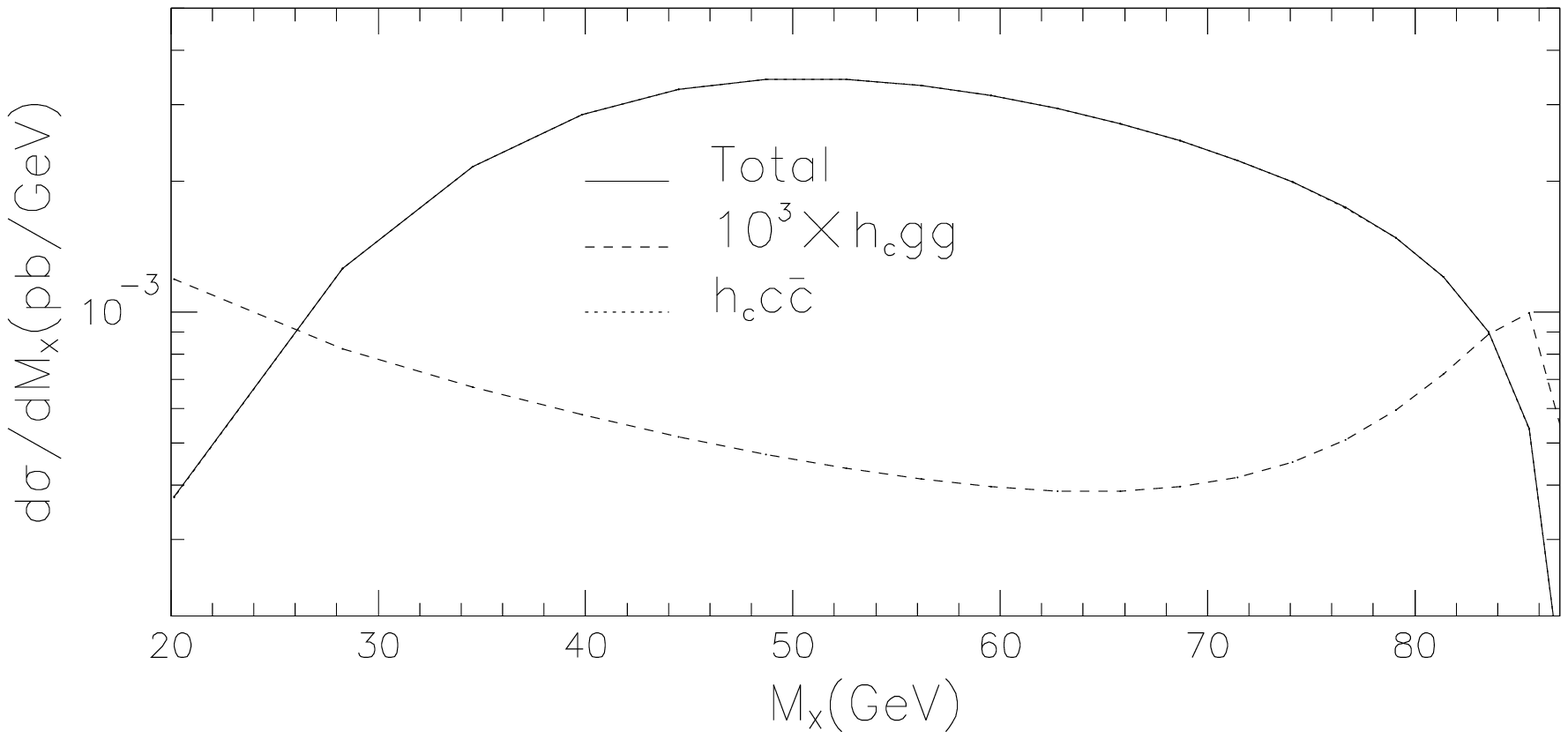}
\caption {\label{fig:gg} distribution of invariant mass of the final states excluding $h_c$ for the process $e^{+}e^{-}\rightarrow h_{c}X$ at the B-factory(upper) and Z-factory(lower).
}}
\end{figure}

\begin{figure}
\center{
\includegraphics*[scale=0.4]{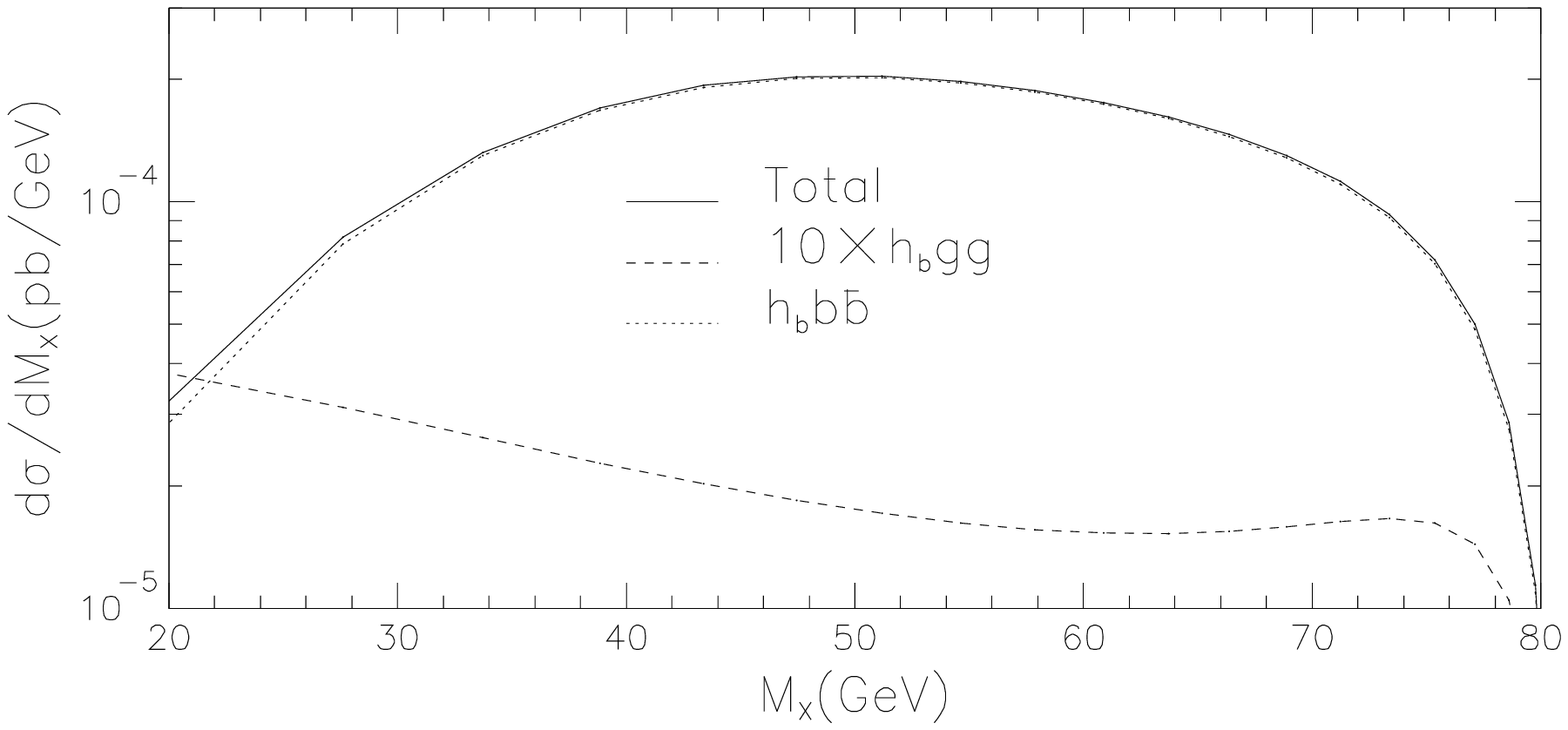}\\
\includegraphics*[scale=0.4]{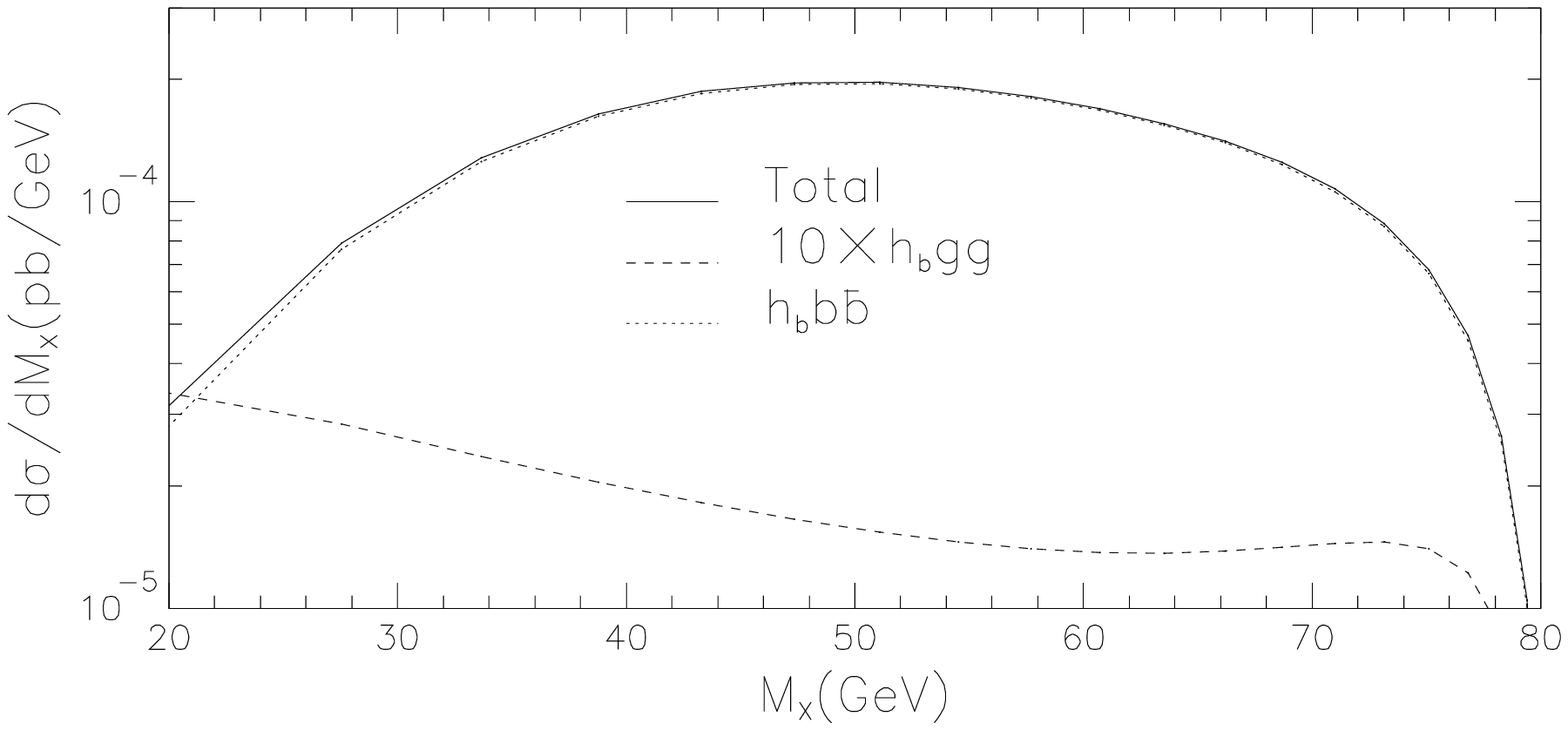}
\caption {\label{fig:ggb} distribution of invariant mass of the final states excluding $h_b$ for the process $e^{+}e^{-}\rightarrow h_{b}X$ for $h_b(1P)$(upper) and $h_b(2P)$(lower) at the Z-factory.
}}
\end{figure}

Fig.~\ref{fig:gg} shows the distribution of invariant mass $M_X$ of the final states with $h_c$ excluded for $h_c$ production 
in the CS processes at B-factory and Z-factory CM energy.
The parameter choice is $m_c=1.76\gev$ and $\alpha_{s}=0.26$.
Fig.~\ref{fig:ggb} gives that for $h_b(1P)$ and $h_b(2P)$ production.
The CM energy is $91.2\gev$ and b-quark mass for $h_b(1P)$ and $h_b(2P)$ are $4.95\gev$ and $5.13\gev$, respectively.
QCD coupling constant is $\alpha_{s}=0.18$.
The $\OP{1}{S}{0}{8}$ process contribute in the phase space point where $M_X$ is zero. 
However, $\OP{1}{S}{0}{8}$ state evolves into a P-wave quarkonium requires it emitting soft gluons 
which will combine with the other gluon and hadronize.
Therefore, its contribution to the $M_X$ distribution is around $M_X=0$, but not exactly at $M_X=0$. 
In the same situation,  the first term in Eq.(\ref{eqn:soft}) 
contribute to the same phase space point of the two body final states $h_c+g$ while the other gluon is soft and has already been integrated out.
It is in the same way to hadronize as the CO process.   
It is obvious that the $M_X$ distribution is not a good observable at small $M_X$ in theoretical calculation.
Although there is no way to do this hadronization and give right distribution since it is a nonperturbative part,
there is a try on this problem for $J/\psi$ production at $e^+e^-$ colliders in Ref~\cite{Fleming:2003gt}.

In summary, we study the inclusive production of $h_c(1P)$, $h_b(1P)$ and $h_b(2P)$ via $e^{+}e^{-}$ 
annihilation with the center-of-mass energy from the CLEO-c energy to $Z^0$ boson mass at leading order of nonrelativistic QCD.  
A detailed discussion on the validity of perturbative calculation for the total cross section of $h_c(h_b)$ production
at the CLEO-c(B-factory) is given. It is easy to see that the perturbative QCD expansion is not good in these cases.
It is also found that validity of the nonrelativistic 
expansion only depends on the value of $v^2$ and has no relation with the momentum of the related gluons, so the nonrelativistic 
expansion is available. Besides, $\mu_\Lambda=m_c$ is not a suitable choice, and $\mu_\Lambda=m_c/2$, which is smaller than the maximum 
gluon energy $1\gev$ in the phase space, is chosen to present the results. 
In our numerical calculation, the LDMEs $\langle O^{h_{c}}(^1S_0^{[8]})\rangle$ and 
$\langle O^{h_{b}}(^1S_0^{[8]})\rangle$ from a fit of the Tevatron experimental data with the LO calculation for 
transverse momentum $p_t$ distribution of $\chi_{cJ}$ hadroproduction are independent on the NRQCD scale 
$\mu_\Lambda$, since the range of $p_t$ in the fit does not include $p_t=0$ where $\mu_\Lambda$ is involved in the 
calculation to factorize IR divergence. Therefore, in our results, $\mu_\Lambda$ independence can not be achieved
and it can bring large uncertainty when the CM energy approach the production threshold. At $Z^0$ boson peak, 
the cross sections are  $39\sim 703$fb for $h_c(1p)$,  $37\sim 61$fb for $h_b(1p)$ and $44\sim 73$fb for $h_b(2p)$. 
At the B-factory, they are  $86\sim212$fb for $h_c(1p)$, $32\sim87$fb for $h_b(1p)$ and $38 \sim 106$fb for $h_b(2p)$ 
where the production rate for $h_b(1p,2p)$ are about 5$\sim$19 times smaller than the experimental measurements. 
And at the CLEO-c, it is $0.3\sim3.1$pb, which is of large uncertainty from the choice of $\mu_\Lambda$, 
far away from the experimental measurement.  In another aspect, the results clearly show that the color-octet
state $(^1S_0^8)$ contributes dominantly while the color-singlet $(^1P_1^1)$ state contributes a small part, in contrast to 
the case of inclusive $J/\psi$ production at the B-factory, where the color-singlet state is found to contribute dominantly.
From previous works at NLO of QCD for inclusive $J/\psi$ production at $e^+e^-$ colliders, it is clear that the QCD 
NLO correction can enhance the production rate $1.7$ times for $e^+e^-\rightarrow J/\psi+cc$~\cite{Zhang:2006ay} and 
$1.2$ times for $e^+e^-\rightarrow J/\psi+gg$\cite{Ma:2008gq}. 
It is quite interesting to see the effect from QCD NLO correction
on the inclusive $h_c(h_b)$ production in future work. 

{\it Note added} When we were preparing our draft, a related work appeared at arXiv recently~\cite{Jia:2012qx}, 
which study on $h_c$ production at B-factories. Once take their renormalization scheme in the calculation of NRQCD 
corrections to CO operator and their parameters, our calculation on $h_c$ production agrees with their numerical results.

This work is supported by the National Natural Science Foundation of
China (Nos.~10979056 and 10935012), in part by DFG and NSFC (CRC 110) and CAS under Project No. INFO-115-B01.


\end{document}